\documentclass{aa} 
\bibpunct{(}{)}{;}{a}{}{,} 
\usepackage{graphicx}
\usepackage{txfonts}
\usepackage{color}

\begin{document}

\title{Dust-driven winds of AGB stars: The critical interplay of atmospheric shocks and luminosity variations}


\author{S. Liljegren
\inst{1}
\and
S. Höfner
\inst{1}
\and
W. Nowotny
\inst{2}
\and
K. Eriksson
\inst{1}
}

\institute{Division of Astronomy and Space Physics, Department of Physics and Astronomy, Uppsala University, Box 516, SE-751 20 Uppsala, Sweden \\
\email{sofie.liljegren@physics.uu.se}
\and
University of Vienna, Department of Astrophysics, Türkenschanzstrasse 17, 1180 Wien, Austria
}

\date{Received ...; accepted ...}

\abstract
{Winds of AGB stars are thought to be driven by a combination of pulsation-induced shock waves and radiation pressure on dust. In dynamic atmosphere and wind models, the stellar pulsation is often simulated by prescribing a simple sinusoidal variation in velocity and luminosity at the inner boundary of the model atmosphere. }
{We experiment with different forms of the luminosity variation in order to assess the effects on the wind velocity and mass-loss rate, when progressing from the simple sinusoidal recipe towards more realistic descriptions. This will also give an indication of how robust the wind properties derived from the dynamic atmosphere models are. }
{Using state-of-the-art dynamical models of C-rich AGB stars, a range of different asymmetric shapes of the luminosity variation and a range of phase shifts of the luminosity variation relative to the radial variation are tested. These tests are performed on two stellar atmosphere models. The first model has dust condensation and, as a consequence, a stellar wind is triggered, while the second model lacks both dust and wind. }
{The first model with dust and stellar wind is very sensitive to moderate changes in the luminosity variation. There is a complex relationship between the luminosity minimum, and dust condensation: changing the phase corresponding to minimum luminosity can either increase or decrease mass-loss rate and wind velocity. The luminosity maximum dominates the radiative pressure on the dust, which in turn, is important for driving the wind. An earlier occurrence of the maximum, with respect to the propagation of the pulsation-induced shock wave, then increases the wind velocity, while a later occurrence leads to a decrease. These effects of changed luminosity variation are coupled with the dust formation. In contrast there is very little change to the structure of the model without dust. }
{Changing the luminosity variation, both by introducing a phase shift and by modifying the shape, influences wind velocity and the mass-loss rate. To improve wind models it would probably be desirable to extract boundary conditions from 3D dynamical interior models or stellar pulsation models. }


\titlerunning{Dust-driven winds of AGB stars}
\authorrunning{S. Liljegren}

\maketitle

%

\section{Introduction}

Atmospheric shock waves, which are triggered by stellar pulsations and large-scale convective motions, play a critical role in the mass-loss mechanism of asymptotic giant branch (AGB) stars. The strong radiating shocks propagate outwards through the stellar atmosphere, intermittently lifting gas to distances where temperatures are low enough to allow for the formation and growth of dust grains. These particles absorb and scatter stellar photons, resulting in an outwards-directed acceleration of the dust. Through dust-gas collisions, momentum is transferred to the gas, triggering a stellar wind. 

Observational support for this scenario comes from several sources. Gas dynamics can be studied quantitatively with high-resolution spectroscopy (line-profile variations), most notably using vibration-rotation lines of CO at infrared wavelengths. The CO molecule plays a special role for atmospheric chemistry; because of the high bond energy it forms deep in the photospheric layers, and is stable throughout the atmosphere and wind acceleration zone. Furthermore, this molecule exists in both C-type and M-type AGB stars. Vibration-rotation lines with different excitation potential, probe different layers in the atmosphere and wind, allowing us to trace the transition from the periodic motions in the atmosphere out to the steady outflows (e.g. \citealt{ADYM1}, \citealt{ADYM2}, \citealt{HINK82}). Line splitting of second overtone CO lines around the luminosity maximum is caused by shock waves passing through the line formation region, giving indications of shock strengths (\citealt{SCHOLZ00}, \citealt{PAP3}). The fundamental mode CO lines show P Cygni profiles which act as a tracer of wind speeds and mass-loss rates. In addition to near-infrared spectra, pure rotational transitions of CO in the radio regime have been used to measure both wind velocities and mass-loss rates (e.g. \citealt{OLOF02}, \citealt{GONZ03}, \citealt{RAM09}). In recent years, interferometry and high-angular resolution imaging have opened up new ways of studying atmospheric dynamics and emerging shock waves (e.g. \citealt{TEJ03}, \citealt{WEIN04}), as well as dust formation, providing important constraints on dust condensation distances for various species (\citealt{KARO13}, \citealt{SACU13}, \citealt{NAKA06}).

One of the most extensively studied feature of AGB stars is the large variation in their luminosity, caused by large amplitude stellar pulsations with substantial radial expansion and contraction of the star. Photometric monitoring in visual and infra-red bands has been conducted for large samples of long period variables (LPVs), giving indications of links between pulsation and mass-loss rates (e.g. \citealt{WHIT}, Fig. 9 of \citealt{ERIK14}). Pulsation triggers sound waves in sub-photospheric layers, which propagate outwards, steepening into strong shocks, and injecting mechanical energy into the upper atmosphere. This creates favourable conditions for efficient dust formation, i.e. high densities in the wake of the shocks at distances where temperatures are low enough to allow for dust formation. In this context, varying luminosity has a twofold effect; it modulates both the atmospheric temperatures and the radiation pressure driving the wind. Dust condensation will preferentially occur around the luminosity minimum as less radiative heating leads to lower temperatures in the atmosphere. On the other hand, stronger radiative acceleration will occur around the luminosity maximum. Therefore mass-loss rates will critically depend on the timing of the shock propagation through the atmosphere, relative to the luminosity variation. In this paper, we use state-of-the-art atmosphere and wind models for carbon stars to study the interplay of luminosity variations and atmospheric shocks, and the consequences for dust formation and mass-loss.

The stellar atmosphere, where radiative effects dominate the structure, represents a regime that is very different from the stellar interior, where convection is the main energy transport mechanism. This is reflected in a tendency of models to focus on one or the other region, i.e. on pulsation and convection or the wind mechanism. Examples of the first type of models are the 1D self-excited pulsation models by \cite{PULS}, or 3D star-in-a-box models by \cite{FREY}. At present, such models have simplified atmospheres and usually do not include wind acceleration. In contrast, time dependent models for dynamic atmospheres and winds of AGB-stars (e.g. \citealt{WIN00}, \citealt{JEO03}, \citealt{DMA3}, \citealt{ERIK14}, \citealt{BLAD15}) typically do not include the deeper layers where stellar pulsation is excited. The inner boundaries of such models are located above the driving zone, just below the stellar photosphere. The effects of pulsations, which are crucial for the dynamics of the atmosphere, are simulated by prescribing a simple temporal variation of luminosity and gas velocities at the inner boundary. This variation usually has a sinusoidal shape, and the phase of the luminosity and gas velocity variation is chosen such that luminosity and radius reach their maximum simultaneously. This approach has the advantage of few free parameters; however, using boundary conditions set in this way does not necessarily reflect a realistic description of AGB star pulsation, as indicated by both models and observations (discussed further in Sect. \ref{22}). Here we experiment with different shapes of the luminosity variations and with phase shifts relative to the sub-photospheric gas velocities,  to study how this affects the structure of the stellar atmosphere, the wind velocity, and the mass-loss rate. Furthermore, based on snapshots of the radial structures, we compute time series of CO vibration-rotation lines to test their diagnostic potential. 

The paper is structured in the following way: Section 2 contains a brief description of the general assumptions and parameters entering the dynamical models, as well as a detailed discussion of the boundary conditions used for luminosity and gas velocity. Section 3 is dedicated to a detailed presentation of the resulting dynamical structures and wind properties. In Section 4 we describe the effects on CO line profiles.

\section{Modelling methods and parameters}

\subsection{Basic ingredients of the dynamical models}
\label{sec21}

To study the interplay of shocks and luminosity variations we use state-of-the-art dynamic atmosphere and wind models of C-rich AGB stars. The models simulate the time-dependent structure and dynamics of the atmosphere and wind, using one-dimensional frequency-dependent radiation-hydrodynamics (RHD) including a detailed description of the dust growth and evaporation. 

The wind-driving dust species in C-stars is amorphous carbon, which grows through reactions involving C, C$_2$, C$_2$H and C$_2$H$_2$. The dust growth and evaporation is treated as a time-dependent process. It is also assumed that there is no drift between the dust and the gas. The system of non-linear partial differential equations, describing gas, dust, and radiation, is solved simultaneously using a Newton-Raphson scheme. The location of the grid points in the models is adaptive, i.e., it takes the density and temperature gradients into account, with the inner boundary placed below the stellar photosphere. The spatial range covered by the model depends on the nature of the resulting atmosphere; if a wind develops, the models outer boundary is at $ \approx 25$ R$_\odot$, while it will be considerably closer to the star for the models with no wind. For a more detailed description of the dynamic models see \cite{DMA3}, \cite{DMA4} and \cite{MAT}.

\subsection{Inner boundary conditions for velocity and luminosity}
\label{22}

\begin{figure}[t]
\centering
\includegraphics[width=\hsize]{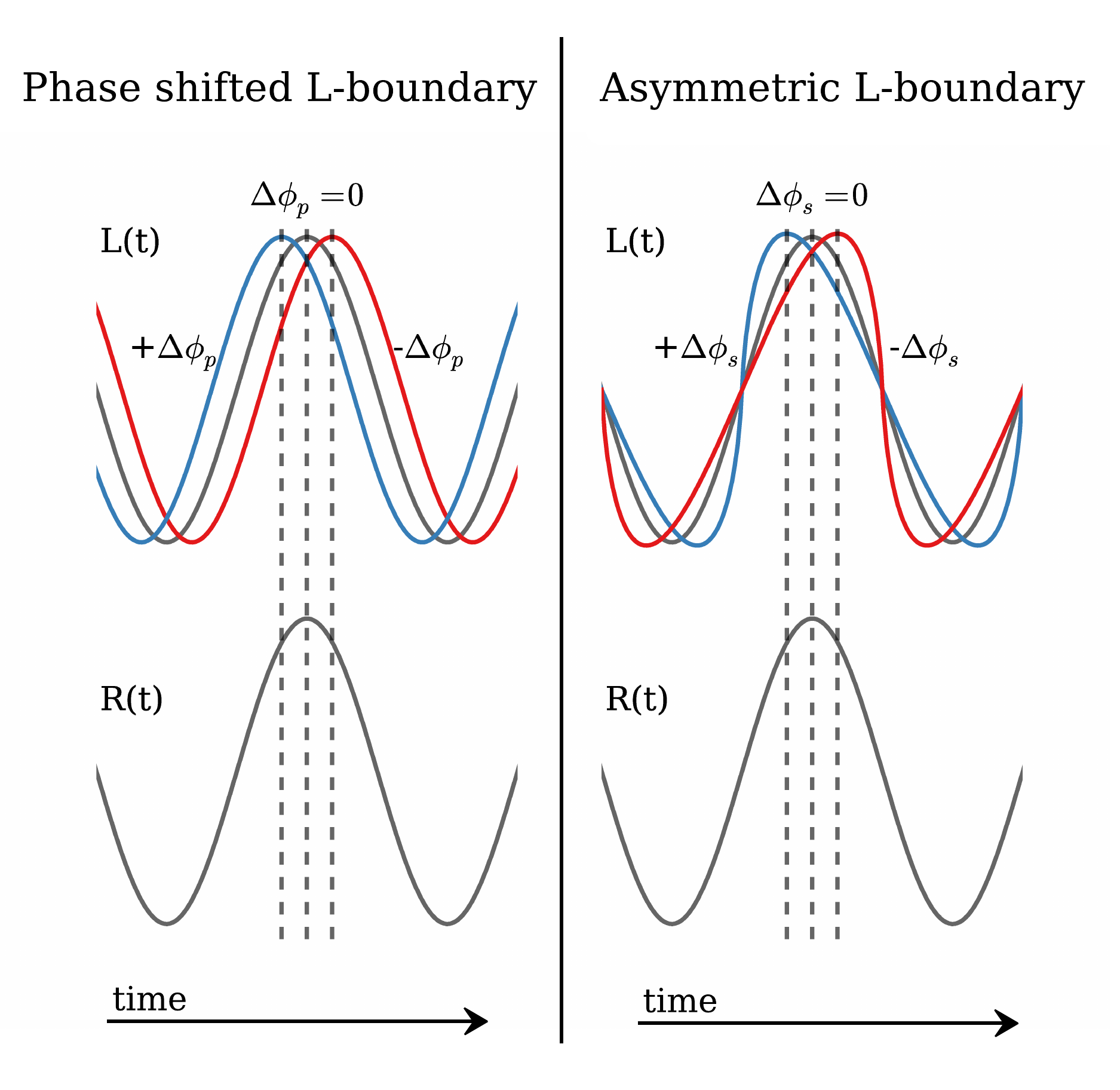}
\caption{Schematic of the two versions of the inner boundary condition used. \textit{Left - }Phase shift $\Delta \phi_p$ between $L(t)$ and $R(t)$. \textit{Right - }Asymmetric luminosity variation, which leads to a phase shift  $\Delta \phi_s$.}
\label{sch}
\end{figure}

As discussed above, the structure of the atmosphere is defined by complex interactions between pulsation, shocks, and radiation pressure on dust. The dynamic models do not include the driving zone of the pulsations, instead the effects are introduced in a parameterised way at the inner boundary. Historically such descriptions have been quite simple, schematically looking like the grey curves in Fig. \ref{sch}. Alternative forms are investigated here. This is done in two ways. Firstly, by introducing a phase shift between the velocity and the luminosity, while keeping the sinusoidal shape, as is the case in the left panel of Fig. \ref{sch} (described in Sect. \ref{221}). And secondly, by making the luminosity variation at the boundary asymmetric, the case in the right panel of Fig. \ref{sch} (described in Sect. \ref{222}). 

The standard approach to simulate the effects of pulsations, introduced by \cite{BOW}, is to define inner boundary conditions where two quantities are varied, $R_{in}(t)$ and $L_{in}(t)$, which correspond to a pulsation of the radius and a variable luminosity input.  A piston at the inner boundary, i.e. below the stellar photosphere, simulates the expansion and contraction of the star as

\begin{equation}
\label{eqn1}
R_{in}(t) = R_0 + \frac{\Delta u_p P}{2 \pi} \sin{\left ( \frac{2 \pi }{P} t \right)}
,\end{equation}

\noindent
The amplitude of the radial movement is dependent on the velocity amplitude $\Delta u_p$ and the period $P$. The lower boundary is impermeable, so there is no gas flow across this layer. The radius variation of Eq.(\ref{eqn1}) therefore corresponds to a gas velocity at the inner boundary of 

\begin{equation}
\label{eqn11}
u_{in}(t) = \Delta u_p \cos\left ( \frac{2 \pi }{P} t \right)
,\end{equation}

The radius variation can be connected to a luminosity variation in a simple way, if constant flux at the inner boundary is assumed. In that case, the luminosity at the inner boundary is proportional to the square of the radius: $L_{in} \propto R^2_{in}(t)$. However, for typical velocity amplitudes used in models, this tends to lead to small bolometric variations. To better match observations of flux variations, a free parameter, $f_L$, was introduced in \citet{DMA4}, to make it possible to adjust the amplitude of the luminosity separately from the radial amplitude, that is to say without changing the mechanical energy input. The form of the luminosity variation is then

\begin{align}
\label{eqn2}
\Delta L_{in}(t) &= L_{in} - L_0= f_L \left (\frac{R^2_{in}(t) - R^2_0}{R^2_0} \right ) \times L_0 \nonumber  \\ 
& =  f_L \left ( \left[1 + \frac{\Delta u_p P}{R_0 2 \pi} \sin{\left ( \frac{2 \pi }{P} t \right)}\right]^2- 1 \right ) \times L_0
\end{align}

\noindent
, where the case $f_L = 1$ corresponds to the case of constant radiative flux at the inner boundary, used in earlier models. Qualitatively this approach describes a stellar interior, which expands and increases in luminosity simultaneously, or similarly compresses and decreases in luminosity simultaneously. 

The method described above was initially introduced because it is simple to implement and it has a small number of free parameters (two in this case, $\Delta u_P$ and $P$). However there are several drawbacks, which motivate the study presented here. A sinusoidal variation of the luminosity is not representative of what is known about AGB stars pulsations (discussed further in Sect. \ref{221} and Sect \ref{222}). It has been suggested that this might cause an inconsistency with the C-star model atmospheres, namely that the loops in a colour-colour diagram (J-H vs H-K), traced out during a pulsation cycle, are not in the same direction as those of observations. (\citealt{SYN2}, \citealt{loop1}, \citealt{loop2}).

There have been attempts to investigate the assumptions about the boundary conditions. \cite{FREY} use 3D interior models to derive non-sinusoidal boundary conditions for the velocity while assuming $L_{in} \propto R^2_{in}(t)$. Their conclusions relevant to this paper were that, while the amplitude of the velocity boundary $u_{in}$ was of great importance for mass-loss rates and wind velocity, the actual shape was not. We are, therefore, not investigating the shape of the velocity boundary condition or the influence of changing the luminosity amplitude. Rather, we look at other pulsation properties, namely the effect of a phase shift between the luminosity variation and the radius variation, and asymmetric shapes of the luminosity variation (see Fig. \ref{sch}).

\subsubsection{Phase shift between radius and luminosity variation}
\label{221}

A phase shift between the radial variation and the luminosity for pulsating AGB stars is something that interior models predict, for example, in \cite{PULS}. The assumption of phase coupling is, therefore, relaxed here, and luminosity and velocity are decoupled at the boundary by introducing a phase shift $\Delta \phi_p$ so that Eq. (\ref{eqn2}) becomes

\begin{equation}
\label{eqn3}
\Delta L_{in}(t, \Delta \phi_p) =  f_L \left (\left[1 + \frac{\Delta u_p P}{2 \pi R_0} \sin{\left ( 2 \pi \left(\frac{t}{P} + \Delta \phi_p \right)  \right)}\right]^2- 1 \right ) \times L_0
,\end{equation}

\noindent

\noindent
while keeping the form of the radial boundary condition $R_{in}$, from Eq.(\ref{eqn1}). The range of phase shifts $\Delta \phi_p \in [0, 1]$ will then span a full pulsation period. This allows us to test the impact of a phase shift between the velocity field, and luminosity (see Fig. \ref{fig1}). 

Although \cite{PULS} predict a positive phase shift, according to this definition, we test a range of values to understand the consequences of decoupling the velocity field and luminosity at the inner boundary, and further investigate if it would be possible to observe such behaviour. 

\subsubsection{Asymmetric luminosity variation}
\label{222}
The other fundamental assumption made in the earlier models about the luminosity variation at the boundary, is that it is symmetric and sinusoidal in shape. Looking at photometry for C-rich LPVs from \cite{WHIT}, and assuming that the K band is representative of the bolometric flux, some light curves show different shapes. \cite{TOMA} further found that around 30$\%$ of Mira light curves varied significantly from a sinusoidal curve, with both different shapes and secondary maxima. This warrants an investigation into the effects of different shapes on luminosity variation.

The effect of having an asymmetric luminosity variation at the inner boundary is tested by representing the luminosity at the inner boundary by a smoothed Fourier sawtooth wave curve as

\begin{equation}
\label{eqn5}
\Delta L_{in}(t) = k \sum \limits_{n=1}^N \frac{1}{n^w} \sin\left ({\frac{2\pi n}{P}t} \right)
,\end{equation}
where $k$ is the amplitude, which is set to match the amplitude of Eq. (\ref{eqn2}) and Eq. (\ref{eqn3}), $P$ is again the period and $w$ is a smoothing factor. Setting $w=1$ leads to a sawtooth wave, however, when increasing the value of $w,$ the resulting $\Delta L$ will be smoothed and approach a symmetric sinusoidal curve. An example of this can be seen in Fig. \ref{sch} or in Fig. \ref{fig2}. 

A measure of how asymmetric the luminosity is, compared to the original boundary condition is introduced by specifying how much the maximum phase of the new boundary is shifted compared to the original boundary. This phase shift is called $\Delta \phi_s$, and is analogous to the phase shift $\Delta \phi_p$, however, as the luminosity is asymmetric now the minimum is shifted in the opposite direction.

\subsection{Model parameters}

\begin{figure}
\centering
\includegraphics[width=\hsize]{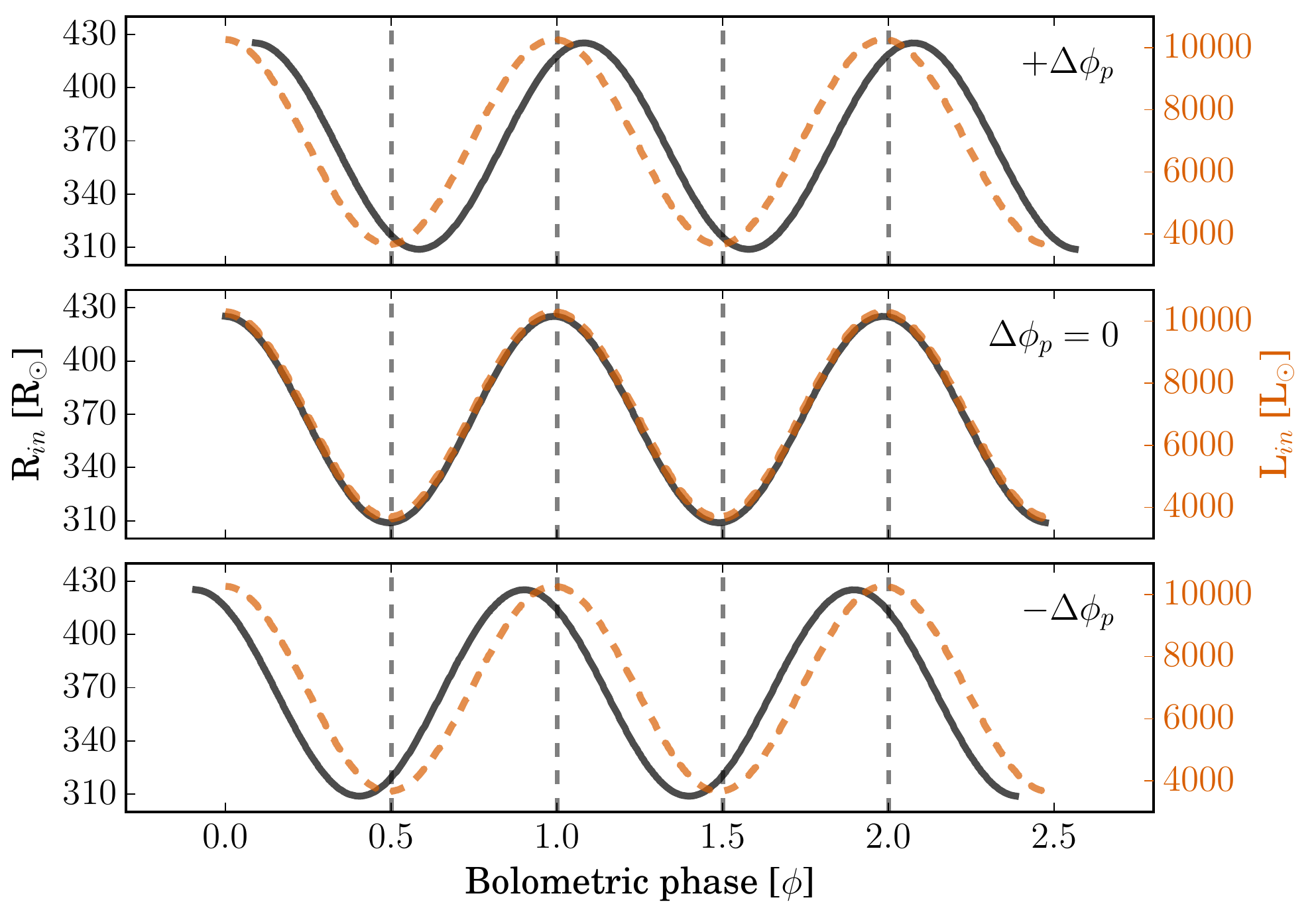}
\caption{Case M-PL. Boundary conditions, R$_{in}$(t) (black, full line) and L$_{in}$(t) (orange, dashed line), for different phase shifts $\Delta \phi_p$. The phase shifts shown are $\Delta \phi_p = \pm 0.09$.
}
\label{fig1}
\end{figure}

\begin{table}[t]
\caption{Parameters of the two dynamical models (adopted from \cite{PAP3}) used in this study. } 
\label{tab1} 
\centering 
\begin{tabular}{l|cc}
\hline
Model: &  W & M \\ \hline
$L_{\star}$ {[}L$_{\odot}${]} & 7000 & 7000 \\
$M_\star$ {[}M$_{\odot}${]} &  1.0 & 1.5 \\
$T_{\star}$ {[}K{]} &  2800 & 2600 \\
$R_{\star}$ {[}R$_\odot${]}  & 355 & 412 \\ \hline
{[Fe/H]} & 0 & 0 \\
C/O {[}by number{]} &  1.4 & 1.4 \\ \hline
Period {[}days{]} &  390 & 490 \\
$\Delta u_p$ {[}kms$^{-1}${]}  & 2 & 6 \\
$f_L$ &  1.0 & 1.5 \\ \hline
\end{tabular}
\tablefoot{While model W represents a pulsating star without mass loss, model M represents a Mira developing a wind. Note that the radius is calculated from luminosity and temperature. }
\end{table}

The dynamical models are characterised by a set of stellar parameters, which are given in Table \ref{tab1}. The models based on these stellar parameters have been used previously by \cite{PAP3}, to examine the connection between line formation and the atmospheric velocity field, where synthetic spectra are compared to high-resolution spectroscopic observations. Similar synthetic spectra are produced here for the models, but with different pulsation properties. The names of the models introduced in the earlier paper, model W and model M, are also used here for continuity.

The dynamical behaviour of the resulting atmospheres comes in two broad categories: one where material is levitated enough to enable dust condensation, which in turn leads to a stellar wind and mass loss, and one where the material is not levitated enough. The models used here represent these two cases (\citealt{PAP3}).

\textit{\textup{Model W} }is a dynamical model with no stellar wind, and therefore no mass loss. The inner boundary of the model pulsates inducing a shock wave once every pulsation cycle. However, the pulsations are too weak to levitate material high enough for dust to condense. The temperatures in the wakes of the shock are never low enough to allow this. Without acceleration from the dust, the layers follow a ballistic trajectory as gravity dominates. 

\textit{\textup{Model M}} represents the other type of AGB atmospheric models with dust condensation and mass loss. The inner dust-free parts of the model behave similarly to model W, with layers following mostly ballistic trajectories. There are, however, large differences in the gas dynamics after dust condensation occurs. With a lower effective temperature, and more pronounced pulsation at the inner boundary, the conditions are more favourable for dust condensing. Once there is dust forming behind a shock wave, radiation pressure acts on the dust grains, resulting in a net outwards momentum, driving stellar wind. Models of such a star become extended and the stellar wind results in significant mass loss.

In both cases, a phase shift between $R_{in}$ and $L_{in}$, and an asymmetric $L_{in}$, are tested for both Model M and Model W. A range of phase shifts, $\Delta \phi_p \in [-0.09, 0.09]$, is investigated. For the asymmetric boundary, values of $w$ in Eq.(\ref{eqn5}) are chosen to mirror the phase shifts of the maxima for the sinusoidal case, with $\Delta \phi_s \in [-0.09, 0.09]$. 

The combination of two models with two boundary conditions leads to four different cases. The shorthand used is given in Table \ref{sh}.

\begin{table}[h]
\centering
\caption{Shorthand for the investigated combinations of model paramaters and boundary conditions. }
\label{sh}
\label{shorthand}
\begin{tabular}{l|c|c}
\hline
                         & Model M & Model W \\ \hline
Phase-shifted luminosity& M-PL    & W-PL    \\
Asymmetric luminosity    & M-AL    & W-AL    \\ \hline
\end{tabular}
\end{table}

\begin{figure}
\centering
\includegraphics[width=\hsize]{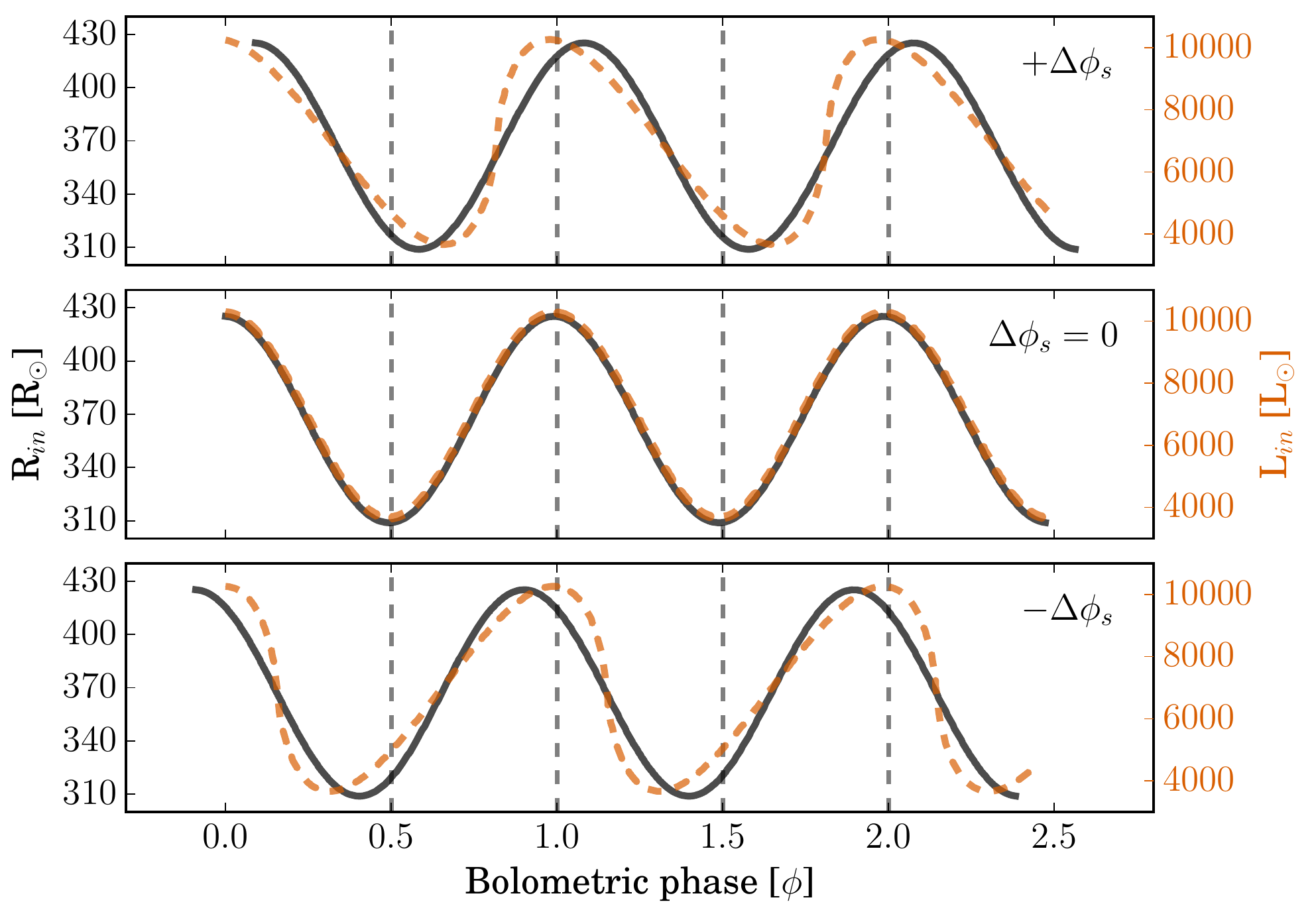}
\caption{Case M-AL. Boundary condition, R$_{in}$(t) (black, full line) and L$_{in}$(t) (orange, dashed line), for an asymmetric boundary where $\Delta \phi_s$ is the difference in bolometric phase between the maximum of R$_{in}$(t) and L$_{in}$(t). The phase shifts shown are $\Delta \phi_s = \pm 0.09$.
}
\label{fig2}
\end{figure}

\section{Atmospheric dynamics and wind properties}

For each of the models M and W, we first describe the behaviour with the original boundary condition, i.e. $R_{in}$ and $L_{in}$ in phase, and then discuss the effects of changing the luminosity at the boundary in the two ways described. The main difference between model M and model W is that model W lacks a wind, as dust never forms in this model. Doing these tests with shifted phase and asymmetric boundary for model W is therefore a useful comparison with model M, as effects due to dust should not be present, but any behaviour caused by other opacity sources should.

\subsection{Model M: Pulsating star with dusty wind}

\begin{figure}
\centering
\includegraphics[width=\hsize]{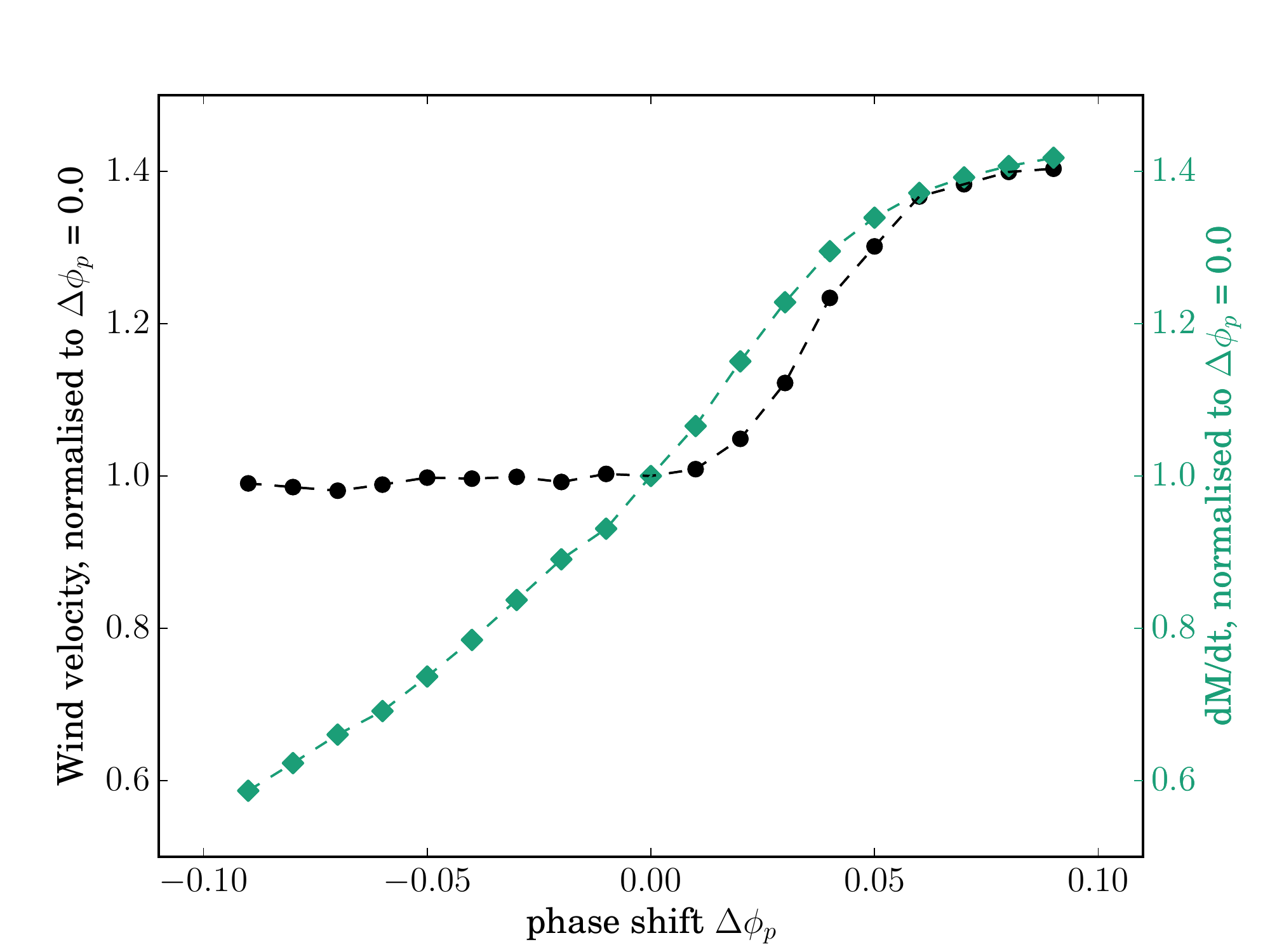}
\caption{Case M-PL. Wind velocity (black circles and left axis) and mass-loss rate (teal-coloured diamonds and right axis) for Model M, normalised to the case $\Delta \phi_p = 0.00$ for phase shift. Wind velocity is 7.7 kms$^{-1}$ and the mass-loss rate is $2.5 \times 10^{-6}$ M$_\odot$yrs$^{-1}$ for $\Delta \phi_p = 0.00$.}
\label{fig3}
\end{figure}

Model M, with unchanged boundary conditions, has typical stellar parameters, and results in a periodic atmosphere and dust-driven outflow with velocity of 7.7 kms$^{-1}$ and a mass-loss rate of $2.5 \times 10^{-6}$ M$_\odot$yrs$^{-1}$. 

The movement of mass shells, the gas temperature, and density as a function of time and radius are shown in the middle row of Fig. \ref{fig4}.

Looking at the mass shell plot (first column) the dust-free part varies with shock waves induced by the piston. Shock waves appear as sudden changes in direction of the curves tracing the movement of the mass shells.

Further out dust is formed in Model M at a regular interval, around $\phi_{bol} = 0.5$. Radiative heating and cooling set the temperatures, seen in the second column, in the relevant layers. The lowest temperature is therefore reached near the luminosity minimum, making this phase favourable for dust condensation. Dust forms most efficiently in the dense material behind the shock wave, and radiation pressure drives the material outwards. This is repeated for every pulsation cycle.

The density structure as a function of time can be seen in column 3 of Fig. \ref{fig4}. Density follows the mass shells, that is to say the movement of the gas, very closely. As seen in the plot, there is a significant increase in density in the shock waves, where outward moving gas encounters slower layers, or gas falling back towards the star. This local compression is superposed onto a general decrease in density outwards.

The picture is different for temperature, seen in column 2. Below 1.5 R$_\star$ the temperature reflects the changing density structure, i.e. the ballistic motions of the hot dust-free gas. Above around 1.5 R$_\star$ molecules, most importantly HCN and C$_2$H$_2$, start to form. These molecules affect the local radiative heating and cooling of the atmosphere, and the temperature structure becomes increasingly complex. At 1.8 R$_\star$ dust condensates and as amorphous carbon dust has a high opacity the radiative pressure will induce a wind. The temperature follows qualitatively the same pattern for every cycle. 

The regular nature of Model M, combined with the fact that it has been previously studied \citep{PAP3}, makes this particular model suitable to investigate what happens to both the wind velocity and the mass-loss rates, as well as the structure of the atmosphere of the star when changing the variation of luminosity at the inner boundary.

\subsubsection{Case M-PL}

A phase shift between the luminosity variation and the radial variation, shown in Fig. \ref{fig1}, results in significant changes in both wind velocity and mass-loss rate. This can be seen in Fig. \ref{fig3}, where these quantities are plotted over a range of phase shifts $\Delta \phi_p$. Wind velocity and mass-loss rate both increase with a positive phase shift, and the two properties appear to be coupled. The wind velocity of the original model is 7.7 kms$^{-1}$ and the mass-loss rate is $2.5 \times 10^{-6}$ M$_\odot$yrs$^{-1}$. The model with the largest phase shift, $\Delta \phi_p = +0.09$, has a wind velocity of 10.8 kms$^{-1}$ and a mass-loss rate of $3.5 \times 10^{-6}$ M$_\odot$yrs$^{-1}$, an increase of approximately $40\%$.

\begin{figure}
\centering
\includegraphics[width=\hsize]{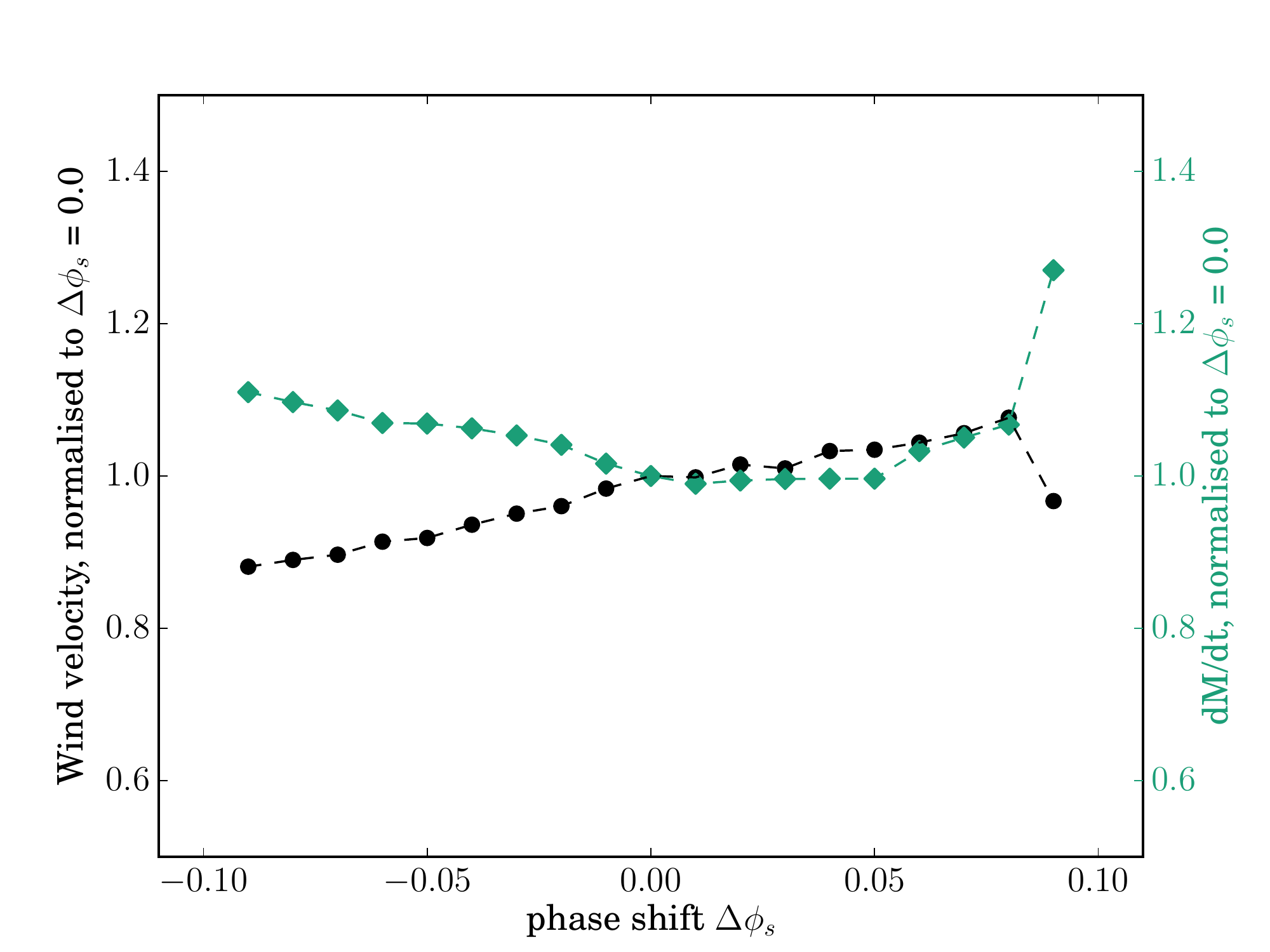}
\caption{Case M-AL. Wind velocity (black circles and left axis) and mass-loss rate (teal-coloured diamonds and right axis) for Model M, normalised for $\Delta \phi_s = 0.00$. Wind velocity is 7.7 kms$^{-1}$ and the mass-loss rate is $2.5 \times 10^{-6}$ M$_\odot$yrs$^{-1}$ for $\Delta \phi_s = 0.00$.}
\label{fig5}
\end{figure}

\begin{figure*}
\centering
\includegraphics[width=17cm]{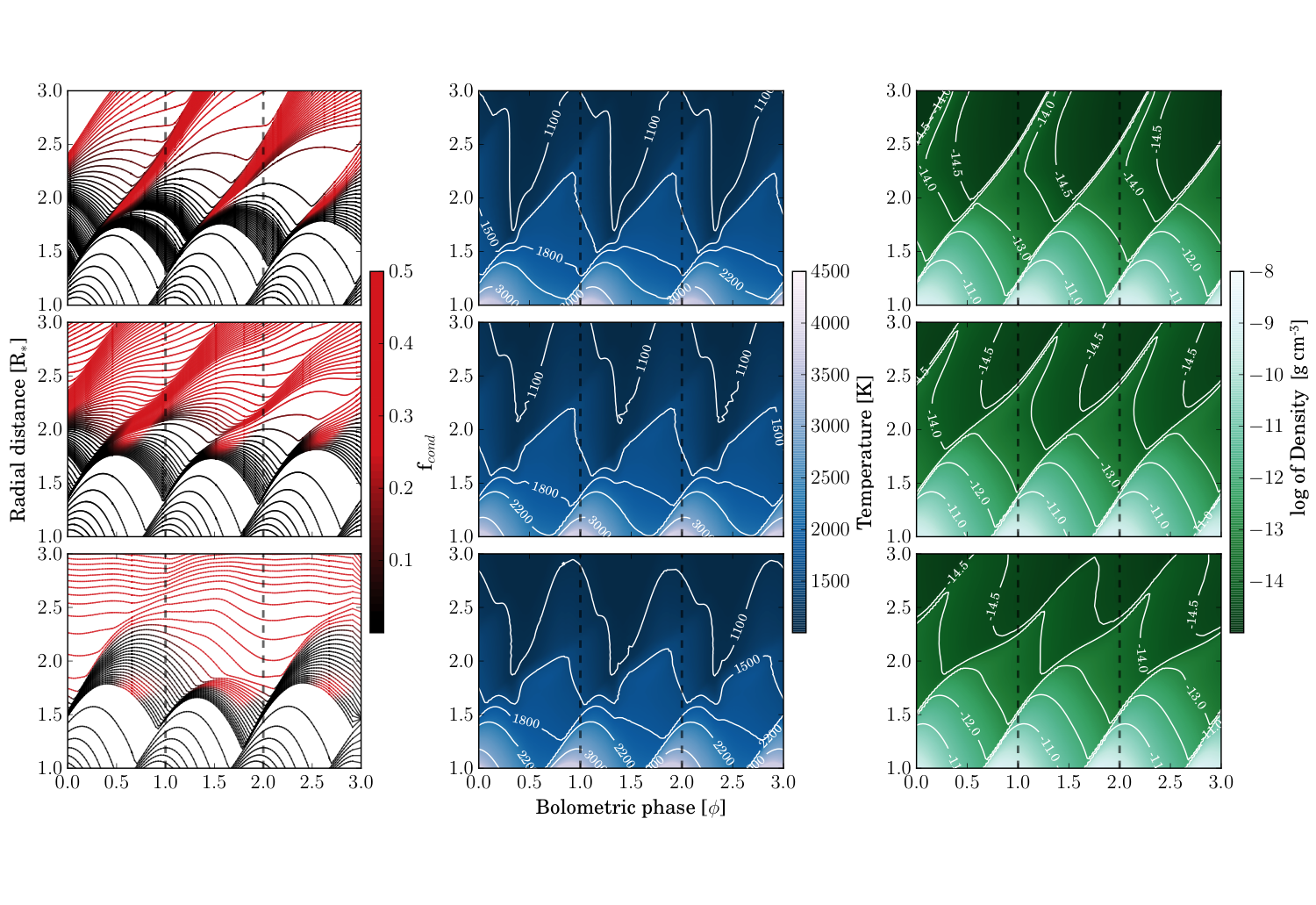}
\caption{Case M-PL. Dynamical structure of Model M for three different phase shifts. \textit{Upper row -} $\Delta \phi_p = +0.09$. \textit{Middle row -} $\Delta \phi_p = 0.00$. \textit{Bottom row -} $\Delta \phi_p = -0.09$. \textit{Left column -} Mass shells with degree of condensation colour-coded. \textit{Middle column -} Gas temperature. The contours are partly picked to indicate where the CO $\Delta v = 2$ lines are formed. \textit{Right column -} Density, with contours showing the logarithm of the gas density.}
\label{fig4}
\end{figure*}

The reason for this can be probed by looking closer at the structure of the atmosphere. The top row of Fig. \ref{fig4} shows the mass shells, the temperature and the pressure structure for the model with largest positive phase shift,  $\Delta \phi_p = +0.09$. There are two effects at play here. Firstly, as the luminosity seems to correlate with temperature, an earlier minimum of the luminosity then leads to lower temperatures earlier, which in turn means that dust is formed earlier in the propagation of the shock wave. This can be seen by comparing the  mass shell plots of $\Delta \phi_p = +0.09$ and $\Delta \phi_p = +0.0$ (mid and top row in Fig. \ref{fig4}). As dust is formed earlier after the shock, the gas layers have a larger velocity when the radiative acceleration due to dust also sets in, increasing the final wind velocity.  Secondly, the maximum of the luminosity occurs earlier during the propagation of the shock wave. A maximum of the luminosity earlier during the pulsation cycle results in larger radiative acceleration of the dust and thus a higher wind velocity. These effects explain the increase in wind velocity with positive phase shift. The mass-loss rate is dependent on both the wind velocity and the density. For positive $\Delta \phi_p = +0.09$ it is mainly the increase in wind velocity which leads to an increase of the mass-loss rate.

For negative phases however, wind velocity and mass-loss rate show different trends. The velocity stays approximately constant while the mass-loss rate continues to decrease, with a decrease of around $40 \%$ to $1.5 \times 10^{-6}$ M$_\odot$yrs$^{-1}$ for $\Delta \phi_p = -0.09$. Fig. \ref{fig4} bottom row shows the case $\Delta \phi_p = -0.09$, and by looking at the mass shell plot the structure is clearly different from the two other cases. In contrast to  $\Delta \phi_p = +0.0$ and  $\Delta \phi_p = +0.09,$ no dust is created directly following the shock wave. Timing of the luminosity minimum relative to shock propagation is unfavourable for dust condensation, as the temperature in the rising material becomes too high in this case. Dust formation then does not occur until later, when material is falling back towards the the star. Some of the dust created collides with the next shock wave, however the amount of material accelerated by this dust is less than when dust is created after the shock wave. The density of the outgoing material is smaller, and while the velocity stays relatively constant, the decrease in density leads to a decrease in mass-loss rate.

What all three cases have in common is that the inner dust-free regions follow approximately ballistic trajectories. Dust condensation takes place around the minimum luminosity phase, which is set by the radiative boundary condition. The differences in where the dust condensation occurs change the morphology of the atmosphere. The density structure follows the shock wave seen in the mass shell plots in Fig. \ref{fig4} , with a steeper shock for $\Delta \phi_p = +0.09$ and a weaker shock wave for $\Delta \phi_p = -0.09$ in the dusty outer layers compared to the original. Similar differences in atmospheric structure are also present for temperature. 

Unfortunately, introducing a phase shift in the boundary does not solve the photometry problem with loops in the wrong direction, previously speculated to be due to simplified pulsation properties. The resulting loops in (J-H vs H-K) are somewhat different in size, probably due to different amounts of dust condensation, however the direction is still wrong \citep{SYN2}.

\subsubsection{Case M-AL}

\begin{figure*}
\centering
\includegraphics[width=17cm]{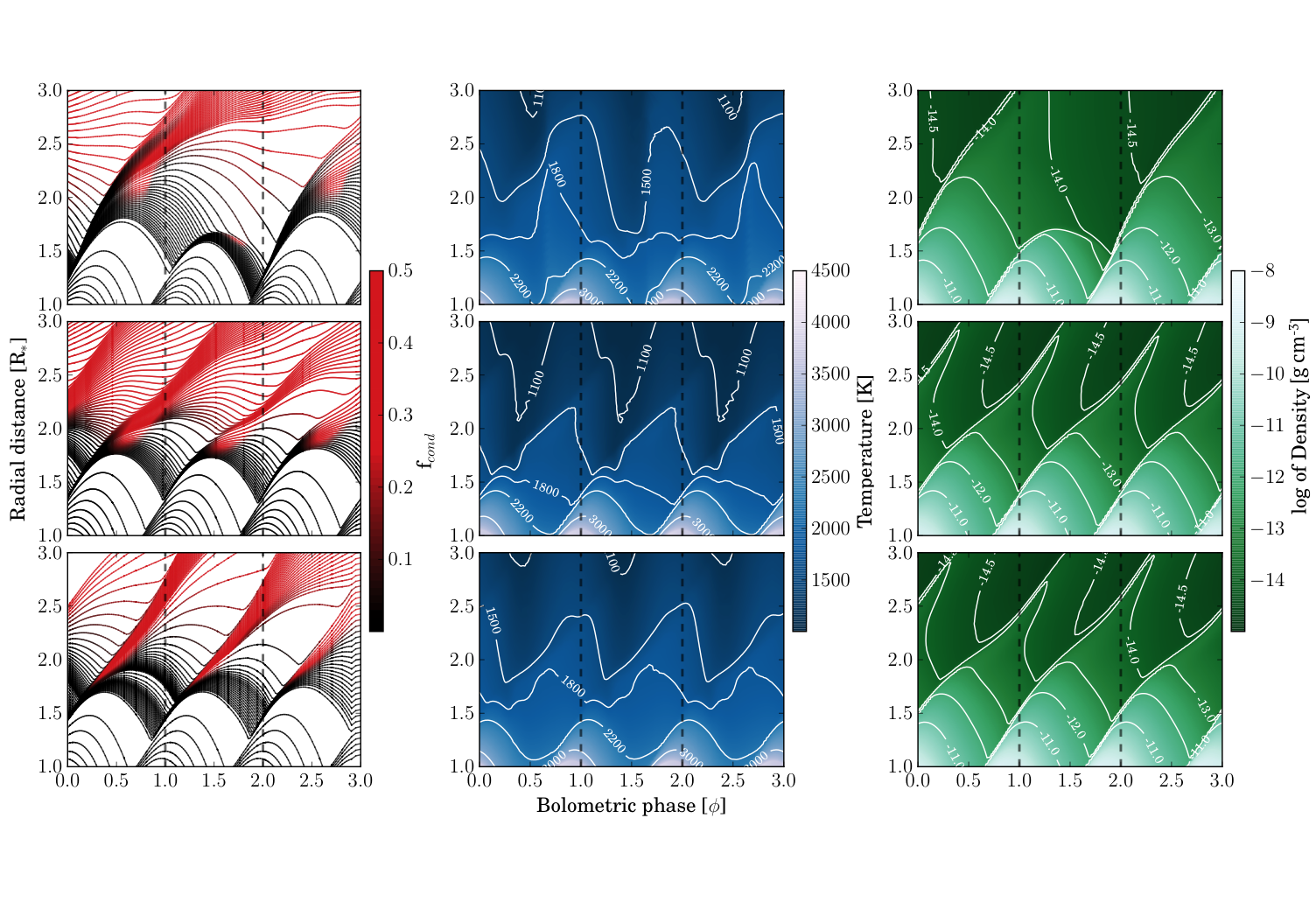}
\caption{Case M-AL. Dynamical structure for Model M for original and two different asymmetric shapes of the luminosity. \textit{Upper row -} $\Delta \phi_s = +0.09$. \textit{Middle row -} $\Delta \phi_s = 0.00$. \textit{Bottom row -} $\Delta \phi_s = -0.09$. \textit{Left column -} Mass shells with degree of condensation colour-coded. \textit{Middle column -} Gas temperature. \textit{Right column -} Density, with contours showing the logarithm of the gas density.}
\label{fig6}
\end{figure*}

An asymmetric variation of the luminosity, as seen in Fig. \ref{fig2}, has effects on the wind velocity and the mass-loss rate. These changes are, however, smaller than for the case M-PL as seen in \ref{fig5}. For increasingly positive $\Delta \phi_s$ there is a small increase in both wind velocity and mass-loss rate, shown in Fig. \ref{fig5}. Changing the boundary asymmetrically this way results in an earlier luminosity maximum in the pulsation cycle, which means larger radiative pressure on the dust and an increase in the wind velocity. This is also analogous to what happened to the maximum of the luminosity in the previous case, M-PL, with positive phase shift. As the luminosity variation is asymmetric however, the minimum will shift in the opposite direction, analogous to the case M-PL with negative phase shift. This means a later minimum with respect to the pulsation cycle, and unfavourable conditions for dust formation. This has a negative effect on the mass-loss rate and wind velocity. The net outcome of the situation is small for the mass-loss rate and wind velocity, as the two effects, in this case the timing of the minimum and the maximum, work in opposite directions. 

The model with $\Delta \phi_s = +0.09$, represented by the data point to the far right, seems to have a peculiar behaviour compared to the models with only slightly less asymmetry. As seen in Fig. \ref{fig6}  the model with $\Delta \phi_s = +0.09$ shows a double periodic behaviour where dust alternatively falls onto the shock wave from the previous shock or forms behind the shock wave. The mass-loss rate increases to $3.1 \times 10^{-6}$ M$_\odot$yrs$^{-1}$, while wind velocity stays about the same as the $\Delta \phi_s =0.0$ model. This type of behaviour has been seen in these models before (discussed in the appendix of \citealt{SYN3}). It is, however, noteworthy that relatively small differences in the boundary condition may change the behaviour of the atmosphere significantly. 

For negative $\Delta \phi_s$ wind velocity and mass-loss rate again follow different trends, in contrast to case M-PL, with an increase in mass-loss rate and a decrease in wind velocity, as shown in Fig. \ref{fig5}. For the case $\Delta \phi_s = -0.09,$ the mass-loss rate increases with around 10$\%$ to $3.0 \times 10^{-6}$ M$_\odot$yrs$^{-1}$. The wind velocity decreases compared to the original model with approximately 10$\%$, to 6.7 kms$^{-1}$. The maximum of the luminosity occurs later during pulsation cycle for negative $\Delta \phi_s$, again similar to how the maximum is shifted by the case M-PL, also leading to a lower wind velocity due to less radiation pressure. The minimum of the luminosity for negative $\Delta \phi_s$ occurs earlier, giving rise to dust condensation early in the shock wave. This leads to a behaviour similar to M-PL, with a positive phase shift, in terms of when and where dust is formed. This can be seen by comparing the mass shell plots of $\Delta \phi_p = +0.09$ in Fig. \ref{fig4}, and $\Delta \phi_s = -0.09$ in Fig. \ref{fig6}. The interpretation of relative phases is thus more complex for an asymmetric boundary, and not perfectly comparable to the previous case, M-PL, where both maximum and minimum are shifted in the same direction and with the same phase difference.

It should further be noted that, similar to the previous case, the inner dust-free part of the atmosphere follows a ballistic trajectory, as seen in Fig. \ref{fig6}. The models differ after the dust is condensed, which again happens around the minimum of the luminosity. Introducing an asymmetric luminosity variation, however, has no effect on the direction of the loops in a (J-H vs H-K) diagram.

\begin{figure}
\centering
\includegraphics[width=\hsize]{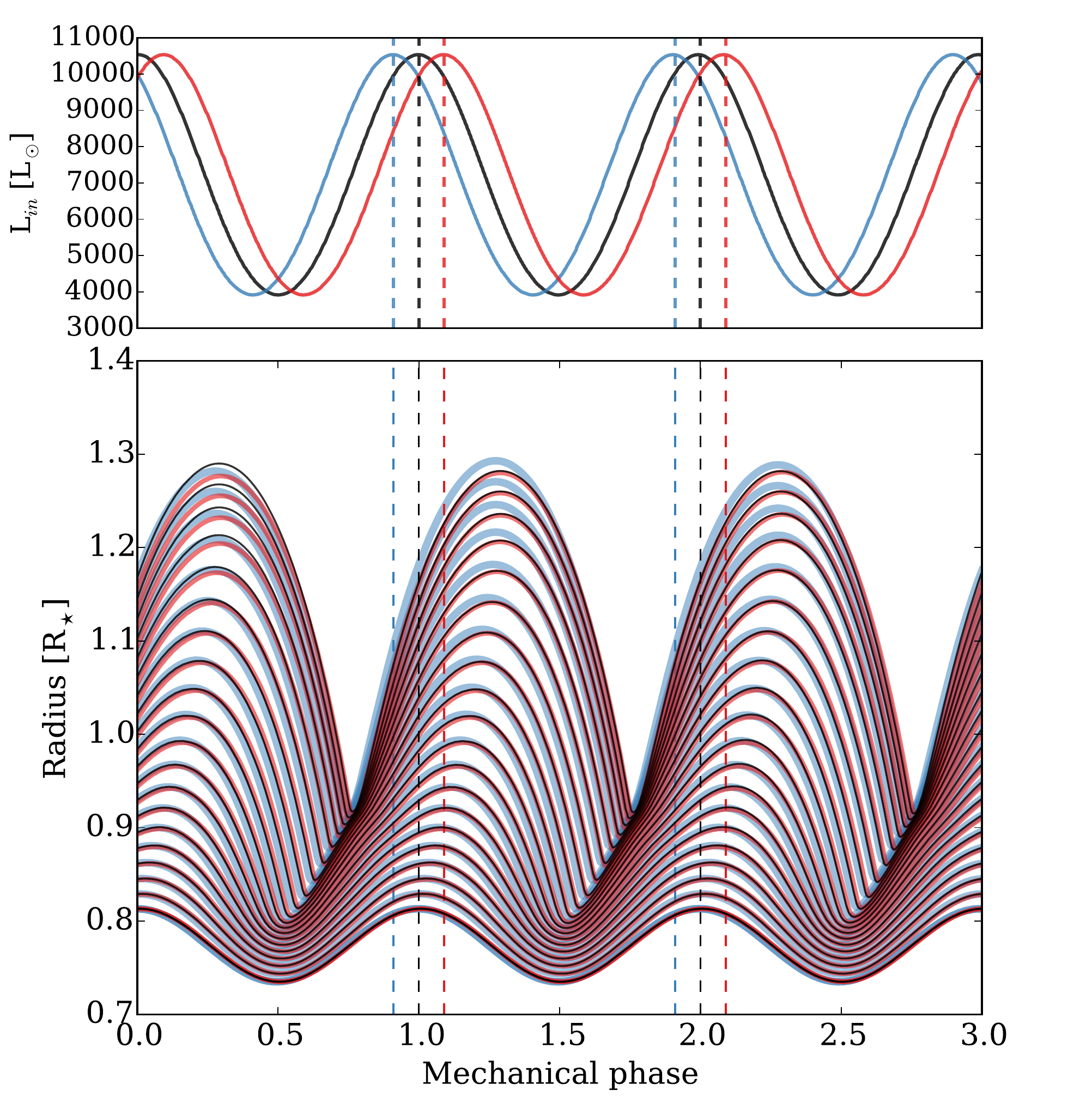}
\caption{Case W-PL. The mass shells of $\Delta \phi_p = +0.09$, $\Delta \phi_p = +0.00$, $\Delta \phi_p = -0.09$ over-plotted, the phase shift removed, where the blue lines correspond to $\Delta \phi_p = +0.09$, the red lines correspond to $\Delta \phi_p = -0.09$ and the black lines to $\Delta \phi_p = 0.00$. Note that mechanical phase is used here, not bolometric phase.  
}
\label{fig71}
\end{figure}
\subsection{Model W: Pulsating star without wind}

Model W has weaker pulsations than Model M. The mass shell plot of Model W is shown in Fig. \ref{fig71}. The black lines correspond to the original boundary condition. The model has regular pulsations, but they are too weak to levitate material to distances where dust can condense. As a result, there is no driving force due to radiation pressure, and the model lacks mass loss.

\subsubsection{Cases W-PL and W-AL}

Lacking dust, which couples luminosity variations to atmospheric dynamics, Model W remains unchanged by the differences in boundary condition. There is essentially no change in the structure of the atmosphere or the velocity field for either phase shift or for asymmetric luminosity. This is illustrated in Fig. \ref{fig71}, where the mass shells for model W with $\Delta \phi_p = +0.09$, $\Delta \phi_p = -0.09$ and $\Delta \phi_p = 0.0$ are plotted overlapping. This is also the case for W-AL, where the corresponding figure is similar to Fig. \ref{fig71}, though it is not shown here. 

This indicates that the radial pulsation of the star dominates this inner region of the atmosphere, while luminosity variation matters little for the structure. The behaviour stays strictly periodic for all cases. Radial excursions of the gas elements are too small to reach sufficiently low temperatures for dust condensation, and as the dust-free gas is optically thin, this model is unaffected by changes in luminosity. The creation of shock waves depends solely on the radial pulsation of the star, which is represented by the velocity boundary condition.

\section{Diagnostics from high-resolution spectra}

\begin{figure}
\centering
\includegraphics[scale = 0.5]{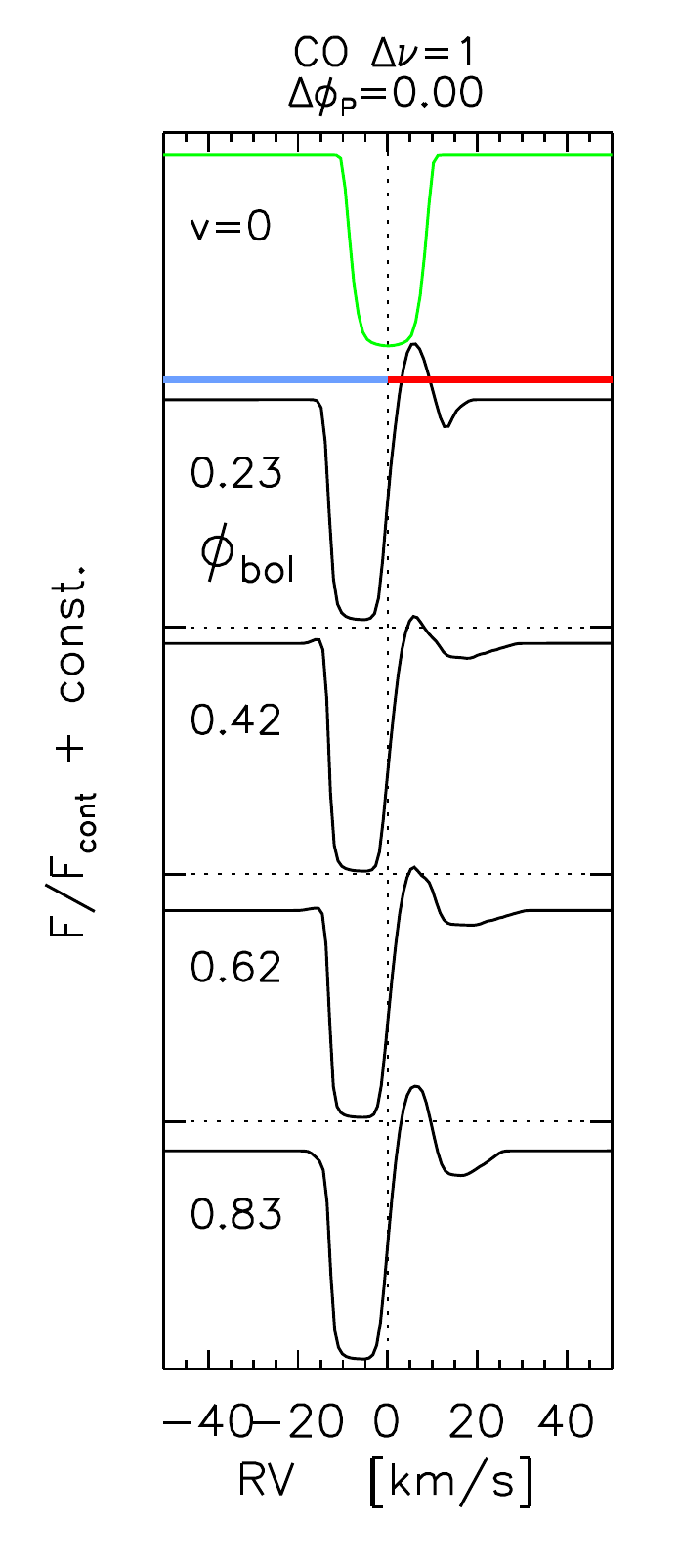}
\caption{CO $\Delta v = 1$ line profiles for the original model M, with $\Delta \phi_p =0$. The line has similar clear P Cygni profiles for all phases. The green line in the upper window is the line without doppler shift, i.e. ignoring gas velocities.
}
\label{pcygni}
\end{figure}

At this point, we ask ourselves if the effects discussed above could be observed. One way to test this is to compare observed light curves with variations of molecular line profiles. Light curves will mostly be dependent on luminosity variation, while molecular line profiles depend on velocity fields and therefore on shock propagation. 

To test the diagnostic potential of line profiles, and to enable comparison with the results by \cite{PAP3}, synthetic spectra are calculated for three CO vibration-rotation lines, whose properties are summarised in Table \ref{tab2}. These lines are of interest because they originate in different regions of the stellar atmosphere, sampling different dynamical behaviour \citep{PAP3}. The temperatures in the line forming regions for CO vibration-rotation lines are roughly 350-500K for CO $\Delta v=1$, 800-1500K for CO $\Delta v=2$ and 2200-3500K for CO $\Delta v=3$. This means that for the chosen lines, CO $\Delta v=1$ probes the outflow, CO $\Delta v=2$ lines will probe mostly the dust-forming regions and CO $\Delta v=3$ lines will probe dust-free inner layers of the pulsating atmosphere. Different aspects of the dynamics can be examined by comparing the behaviour of these lines (see \cite{PAP3} for a detailed discussion).

The synthetic line profiles are computed in the following way: frequency dependent opacities are produced using the COMA code, with assumptions of LTE, micro-turbulence of $\xi = 2.5$kms$^{-1}$ and the line shapes are described by Doppler profiles. Other opacity sources are ignored to clearly discern the differences in the lines caused by different boundary conditions. Radiative transfer is solved for the resulting opacities, using the code by \cite{RADTRAN}. Because of the relatively wide line-forming regions in AGB atmospheres, sphericity must be taken into account. Infall and outflow of gas will define the line profiles so the influence of the gas velocity has to be accounted for in the radiative transfer. For an in-depth description of how the spectral synthesis is performed see \cite{ADYM1, ADYM2, PAP3}. Synthetic spectra for the three lines described were produced for a series of snapshots of Model M, using the extreme models $\Delta \phi_p = -0.09, 0.0, +0.09$ for phase shift and $\Delta \phi_s = -0.09, 0.0, +0.09$ for asymmetric shape.

\begin{figure}
\centering
\includegraphics[scale = 0.42]{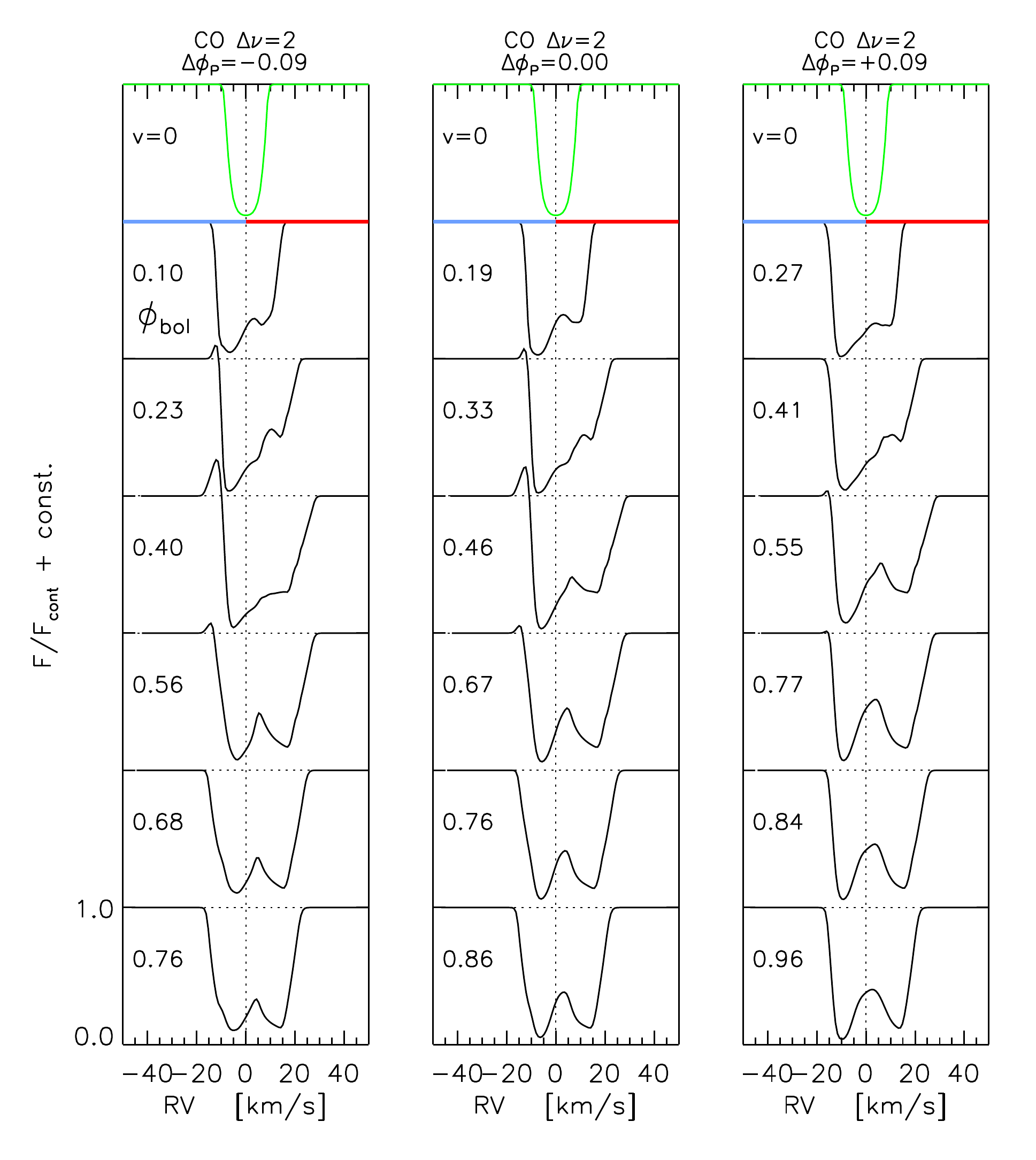}
\caption{Line profiles for CO $\Delta v = 2$ line for case M-PL. The bolometric phases (indicated in each panel) are picked to highlight interesting features.\textit{Left -} $\Delta \phi_p = -0.09$, \textit{Middle -} $\Delta \phi_p = 0.00$, \textit{Right -} $\Delta \phi_p = -0.09$
}
\label{figline1}
\end{figure}

\label{line}
\begin{table}[t]
\caption{Properties of the CO vibration-rotation lines used for the spectral synthesis.} 
\label{tab2} 
\centering 
\begin{tabular}{ll|cc}
\hline
Line & Designation &  $wn${[}cm$^{-1}${]} & $\lambda$ {[}$\mu$m{]} \\ \hline
CO $\Delta v = 1$ & 1-0 R1 & 2150.8560 & 4.6493 \\
CO $\Delta v = 2$ & 2-0 R19  & 4322.0657 & 2.3137 \\
CO $\Delta v = 3$ & 5-2 P30  & 6033.8967 & 1.6573 \\ \hline
\end{tabular}
\tablefoot{These specific lines were adopted from \cite{ADYM1, PAP3}.}
\end{table}

The fundamental mode vibration-rotation lines of CO, with the CO 1-0 R1 line specifically studied here, originate in layers with relatively low gas temperatures, and therefore sample the wind region with constant outflow of the AGB star. Gas velocity and flow rate at this distance, around $R \approx 15 R_\star$, are essentially constant, and the resulting line shapes are P-Cygni profiles, with a blue absorption component from the outflow in the line-of-sight material as well as a red emission component from the extended envelope. The P-Cygni profiles are present for all cases and all phases, with little variation. An example of these can be seen in Fig. \ref{pcygni}. The shape of the line is the same for all phases, with small variations in strength due to changes in velocity. This indicates negligible diagnostic potential for the fundamental mode lines, in regards to pulsation and shock propagation. They can, however, be used to deduce mass-loss rates and wind velocities. Note that all cases of different boundary conditions will look qualitatively similar to that of Fig. \ref{pcygni}.

\begin{figure}
\centering
\includegraphics[scale = 0.42]{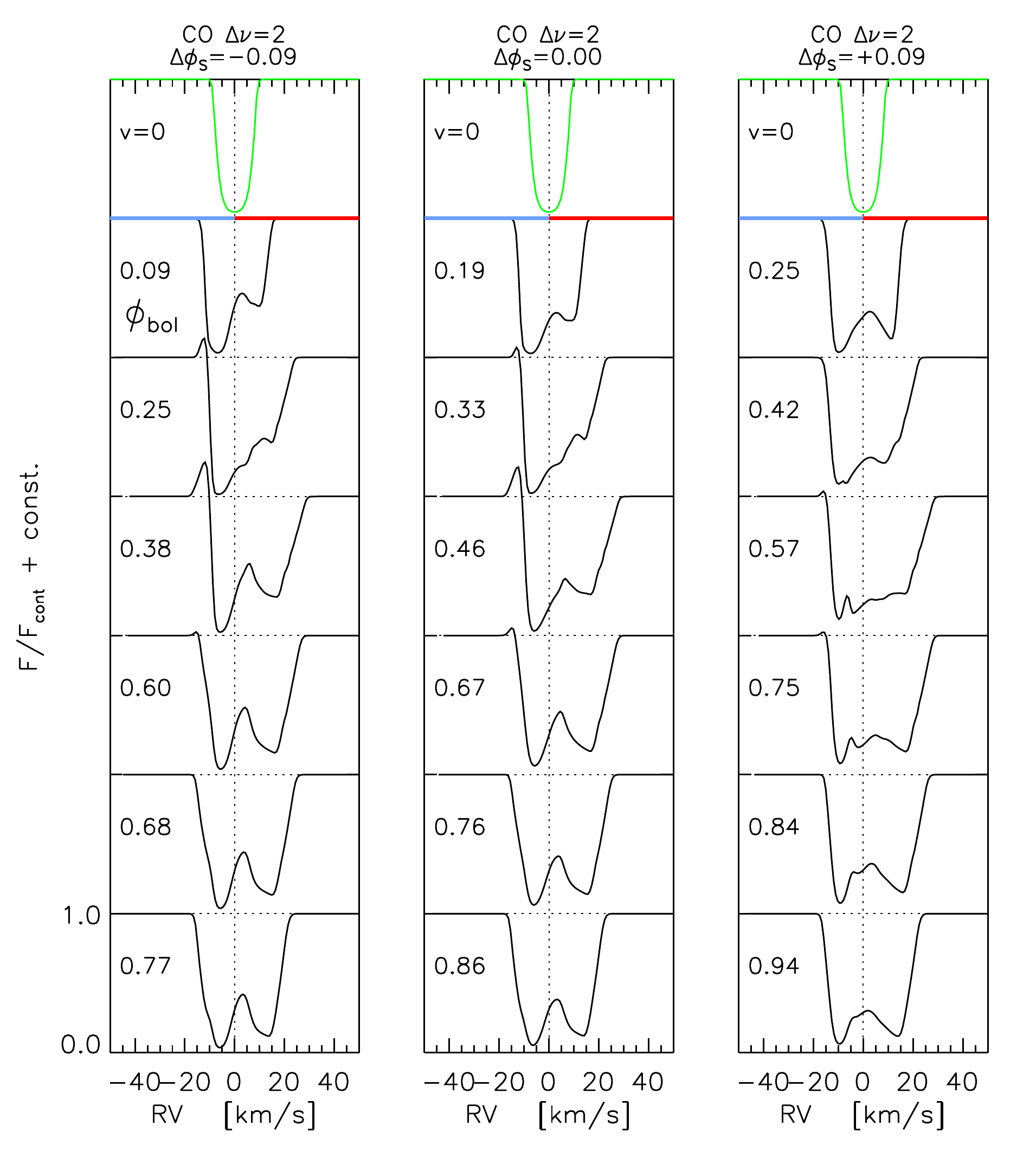}
\caption{Line profiles for CO $\Delta v = 2$ line for case M-AL. \textit{Left -} $\Delta \phi_s = -0.09$, \textit{Middle -} $\Delta \phi_s = 0.00$, \textit{Right -} $\Delta \phi_s = -0.09$}
\label{figline2}
\end{figure}

The line-forming regions for the first and second overtone lines are both in regions of the atmosphere where the effects of stellar pulsation are more prominent, with both out-flowing and in-falling material. The first's overtone CO lines, here represented by CO $\Delta v=2$ 2-0 R19, originate at around $R \approx 2 R_\star$, which are the layers where dust is being formed. The shapes of the line are complex, asymmetric and time dependent, which is true for all cases, as seen in Figs. \ref{figline1} and \ref{figline2}. Qualitatively, the shapes are similar to results found in \cite{ADYM1} and \cite{PAP3}, however there are some small differences between the models with different boundary conditions, probably due to the difference in dust condensation and resulting acceleration. Similar line profiles will occur at different bolometric phases; earlier for negative phase shift, and later for positive phase shift (see phases indicated in the panels of Fig. \ref{figline1} and Fig. \ref{figline2}). This makes sense as the features reflect gas movement, which is shifted in bolometric phase, depending on the boundary condition. 

\begin{figure}
\centering
\includegraphics[scale = 0.42]{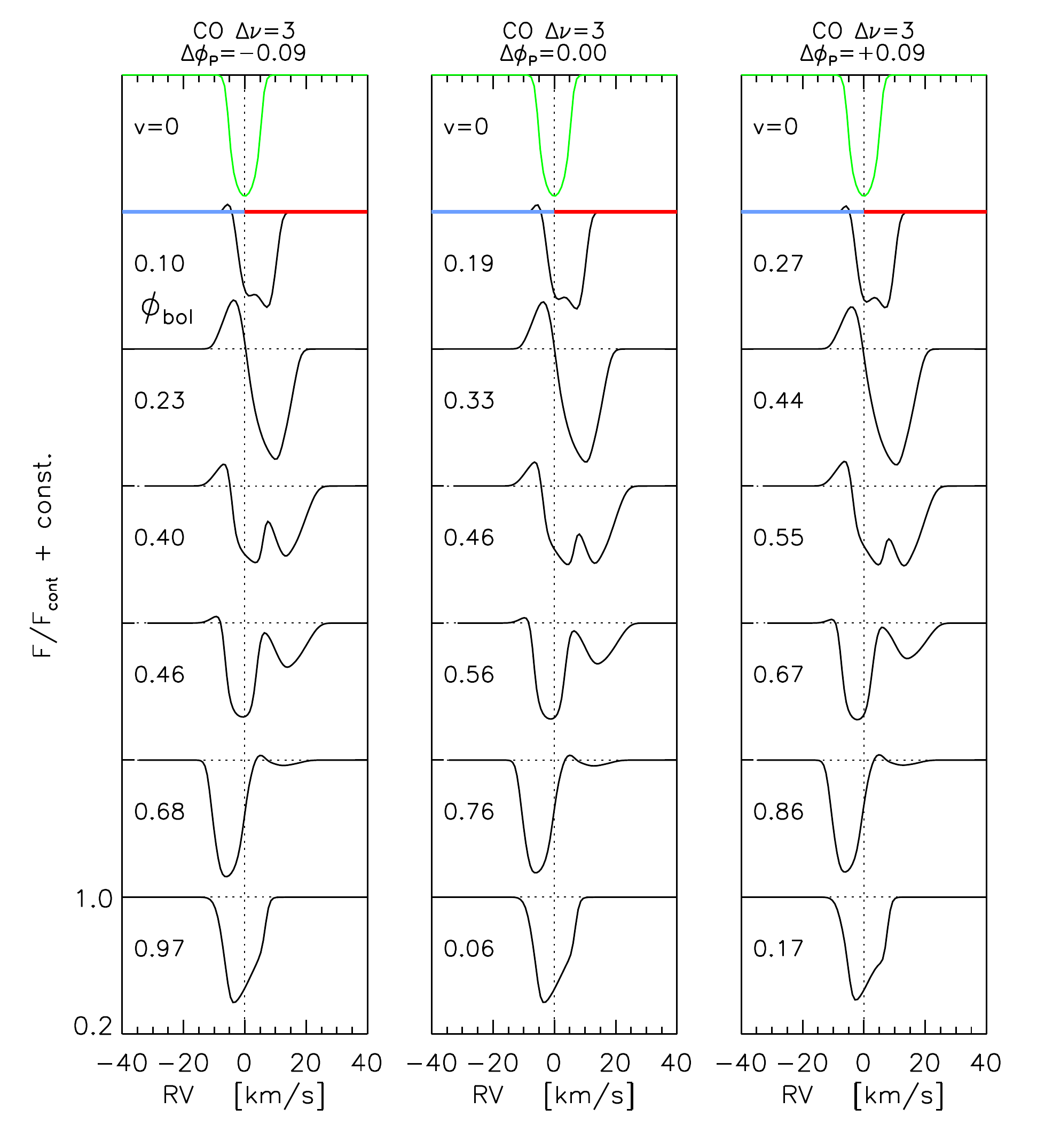}
\caption{Line profiles for CO $\Delta v = 3$ line for case M-PL. The phases are picked to highlight interesting features. \textit{Left -} $\Delta \phi_p = -0.09$, \textit{Middle -} $\Delta \phi_p = 0.00$, \textit{Right -} $\Delta \phi_p = -0.09$
}
\label{figline3}
\end{figure}

\begin{figure}
\centering
\includegraphics[scale = 0.42]{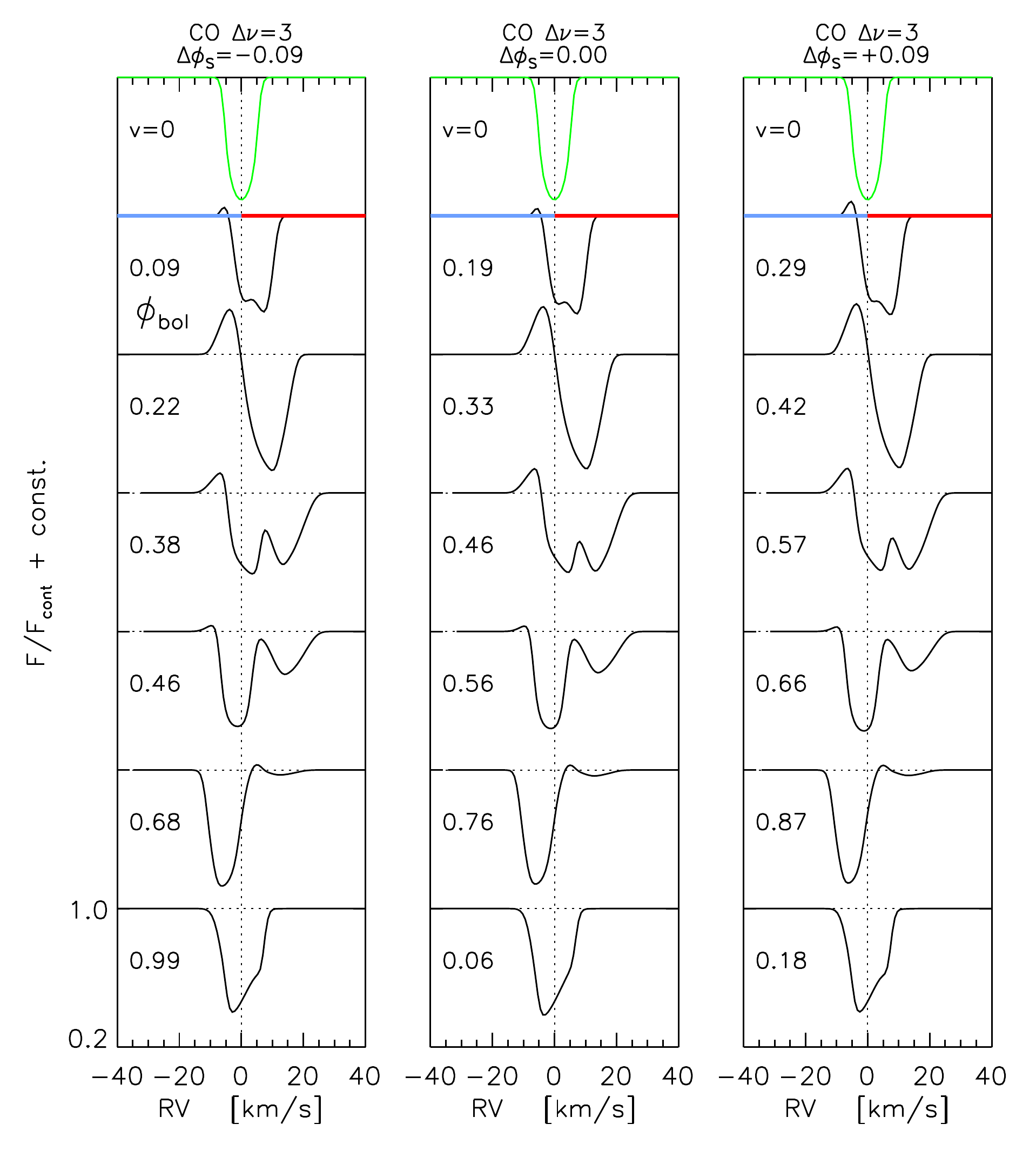}
\caption{Line profiles for CO $\Delta v = 3$ line for case M-AL. \textit{Left -} $\Delta \phi_s = -0.09$, \textit{Middle -} $\Delta \phi_s = 0.00$, \textit{Right -} $\Delta \phi_s = -0.09$}
\label{figline4}
\end{figure}

This is even more obvious for the second overtone CO line tested, CO $\Delta v=3$ 5-2 P30, seen in Figs. \ref{figline3} and \ref{figline4}. This line originates below the dust-forming layers. The lines show time dependent behaviour, with red-shifted components at phases where material is in-falling, blue-shifted when material out-flowing and line doubling at times when shock waves pass through the region of line formation. Fig. 1 of \cite{TOM1} shows a very good schematic of this scenario. This behaviour occurs for all models, but for different bolometric phases. Results are qualitatively similar when comparing features, bar phase shift. 

Again, the fact that the CO $\Delta v=3$ profiles for the three models do not differ in shape (as CO $\Delta v=2$), supports the conclusion that the inner, dust-free part of the atmosphere is not sensitive to changes in the luminosity at the boundary, i.e. is not subject to significant radiation pressure. This part of the atmosphere is dominated by radial expansion and compression of the star, which induces shock waves.  It makes observations of the CO $\Delta v = 3$ lines suitable for deducing information about shock propagation through the atmosphere as consequence of stellar pulsation. The line features seen depend on the shock waves, which in turn depend on the gas movement. Observational information about possible phase shifts between the luminosity and radial variation could then be extracted by comparing line data, which will give insight into the radial contraction and expansion, with light curves, which indicate the bolometric phase. This information could potentially be used as a diagnostic tool for pulsation models.

\section{Summary and Conclusions}

To summarise the results in this paper:

\begin{itemize}
\item The dynamics of the inner, dust-free region of the atmosphere are not sensitive to changes in the luminosity. Gas opacities are too low to produce significant radiative acceleration.
\item Dust condensation however, is very sensitive to changes in luminosity, due to its effects on temperature. Dust condensation sets in at approximately the luminosity minimum, when less radiative heating leads to lower temperatures.
\item If the luminosity minimum occurs early in the pulsation cycle, the density in the condensation zone is higher leading to more efficient dust formation. This results in a larger mass-loss rate. However, if the dust forms late, because of the luminosity minimum occurring later stage, less material is accelerated and the mass-loss rate is reduced. The timing between dust formation, which is set by the luminosity variation, and the propagation of the shock wave, set by the radial expansion and contraction of the star, is then very important for both mass-loss rate and wind velocity, as well as the structure of the atmosphere.
\item The maximum luminosity is important for wind velocity. If the maximum occurs early in comparison to the propagation of the shock wave, there is more radiative pressure on the dust, resulting in a higher wind velocity. If the maximum occurs late, the wind velocity is lower. 
\item The effects of a phase lag between variation in  luminosity and the radius might be possible to observe. The shape of the CO $\Delta v = 3$ line is determined by the propagation of the shock wave, which in turn depends primarily on the radius variation due to pulsation. Consequently, line profile variations, combined with information about the bolometric phase, can be used as a diagnostic tool.
\item One motivation for the current study was a known discrepancy between the photometry of the dynamical wind models and observations. In \cite{SYN2}, it was found that the models loop through the (J-H)-(H-K)-plane in the opposite direction during a light cycle compared with the observed target RU Vir.  While the resulting colour-colour diagram loops are somewhat consistent with observed ones in size and shape, both for the old and the new pulsation properties, the direction does not change for any of the models tested here. This suggests that the inner boundary condition is not the reason for this effect.
\end{itemize}

\noindent
The simplified description used in earlier models to represent stellar pulsation, with both radial variation and luminosity variation, described by sinusoidal curves locked in phase at the inner boundary of the atmospheric models, plays a role in the resulting mass-loss rate and wind velocity. There are indications that more physical boundary conditions might include both phase shifts and different shapes (e.g. \citealt{WHIT}, \citealt{PULS}). Moderate changes to shape and phase for typical stellar parameters lead to changes in wind velocity and mass-loss rate of the the order $\pm 40 \%$. 

To make the dynamical model atmospheres more realistic, suitable boundary conditions should be extracted from pulsation models, both 1D and 3D. Synthetic spectra produced could then be compared with observations, making this a possible diagnostic tool for interior models, as line profile variations reflect the pulsation properties of the star.

\begin{acknowledgements}
SH acknowledges support from the Swedish Research Council (Vetenskapsrådet). 
\end{acknowledgements}


\bibliographystyle{aa} 
\bibliography{ref.bib} 
\end{document}